\documentclass[10pt,twocolumn,letterpaper]{article}
\usepackage[accsupp]{axessibility}
\usepackage{iccv}
\usepackage{times}
\usepackage{epsfig}
\usepackage{amsmath}
\usepackage{amssymb}

\usepackage{cite}
\usepackage{float}
\usepackage{color}
\usepackage{enumerate}
\usepackage{graphics}
\usepackage{graphicx}
\usepackage{adjustbox}
\usepackage{xspace}
\usepackage{url}
\usepackage{booktabs}
\usepackage{array}
\usepackage{textcomp}
\usepackage{rotating}
\usepackage{multirow}
\usepackage{soul}
\makeatletter
\@namedef{ver@everyshi.sty}{}
\makeatother
\usepackage{svg}

\usepackage{tabularx}
\newcolumntype{L}{>{\centering\arraybackslash}m{1.5cm}}
\newcolumntype{S}{>{\centering\arraybackslash}m{2.0cm}}
\newcolumntype{s}{>{\centering\arraybackslash}m{1.0cm}}
\newcolumntype{H}{>{\setbox0=\hbox\bgroup}c<{\egroup}@{}}

\usepackage{caption}
\usepackage{subcaption}
\usepackage{outlines}
\usepackage{makecell}
\usepackage{authblk}

\newif\ifmarkedup
\markeduptrue

\ifmarkedup
    
\fi
\usepackage[breaklinks=true,bookmarks=false]{hyperref}

\iccvfinalcopy 

\def\iccvPaperID{23} 

\ificcvfinal\fi
\begin{document}

\title{A Pathology Deep Learning System Capable of Triage of Melanoma Specimens Utilizing Dermatopathologist Consensus as Ground Truth \thanks{Acknowledgements: The authors thank the support of Jeff Baatz and Liren Zhu at Proscia for their engineering support; Theresa Feeser, Pratik Patel, and Aysegul Ergin Sutcu at Proscia for their data acquisition and Q\&A support; and Dr. Curtis Thompson at CTA and Dr. David Terrano at Bethesda Dermatology Laboratory for their consensus annotation support. }
}
\author[1]{Sivaramakrishnan Sankarapandian}
\author[1]{Saul Kohn}
\author[1]{Vaughn Spurrier}
\author[1]{Sean Grullon}
\author[1]{Rajath E. Soans}
\author[1]{Kameswari D. Ayyagari}
\author[1]{Ramachandra V. Chamarthi}
\author[2]{Kiran Motaparthi}
\author[3]{Jason B. Lee}
\author[4]{Wonwoo Shon}
\author[1]{Michael Bonham}
\author[1]{Julianna D. Ianni}
\affil[1]{Proscia Inc.}
\affil[2]{Department of Dermatology, University of Florida College of Medicine}
\affil[3]{Department of Dermatology, Sidney Kimmel Medical College at Thomas Jefferson University}
\affil[4]{Department of Pathology and Laboratory Medicine, Cedars-Sinai}

\maketitle
\ificcvfinal\thispagestyle{empty}\fi

\begin{abstract}
Although melanoma occurs more rarely than several other skin cancers, patients' long term survival rate is extremely low if the diagnosis is missed. Diagnosis is complicated by a high discordance rate among pathologists when distinguishing between melanoma and benign melanocytic lesions.  A tool that allows pathology labs to sort and prioritize melanoma cases in their workflow could  improve  turnaround  time by prioritizing challenging cases and routing them directly to the appropriate subspecialist. We present a pathology deep learning system (PDLS) that performs hierarchical classification of digitized whole slide image (WSI) specimens into six classes defined by their morphological characteristics, including classification of ``Melanocytic Suspect'' specimens likely representing melanoma or severe dysplastic nevi. We trained the system on 7,685 images from a single lab (the reference lab), including the the largest set of triple-concordant melanocytic specimens compiled to date, and tested the system on 5,099 images from two distinct validation labs. We achieved Area Underneath the ROC Curve (AUC) values of 0.93  classifying Melanocytic Suspect specimens on the reference lab, 0.95 on the first validation lab, and 0.82 on the second validation lab. We demonstrate that the PDLS is capable of automatically sorting and triaging skin specimens with high sensitivity to Melanocytic Suspect cases and that a pathologist would only need between 30\% and 60\% of the caseload to address \emph{all} melanoma specimens.
\end{abstract}

\section{Introduction}

More than 5 million diagnoses of skin cancer are made each year in the United States, about 106,000 of which are melanoma of the skin \cite{acs2021stats}. Diagnosis requires microscopic examination of hematoxylin and eosin (H\&E) stained, paraffin wax embedded biopsies of skin lesion specimens on glass slides. These slides can be manually observed under a microscope, or digitally on a whole slide image (WSI) scanned on specialty hardware. 

The 5-year survival rate of patients with metastatic malignant melanoma is less than 20\% \cite{noone2017cancer}. Melanoma occurs more rarely than several other types of skin cancer, and its diagnosis is challenging, as evidenced by a high discordance rate among pathologists when distinguishing between melanoma and benign melanocytic lesions ($\sim40\%$ discordance rate; e.g. \cite{gerami2014histomorphologic, elmore2017pathologists}). The Melanocytic Pathology Assessment Tool and Hierarchy for Diagnosis (MPATH-Dx; MPATH hereafter) reporting schema was introduced by Piepkorn \textit{et al.} \cite{piepkorn2014mpath} to provide a precise and consistent framework for dermatopathologists to grade the severity of melanocytic proliferation in a specimen. MPATH scores are enumerated from I to V, with I denoting a benign melanocytic lesion and V denoting invasive melanoma. It has been shown that discordance rates are related to the MPATH score, with better inter-observer agreement on both ends of the scale than in the middle (e.g. \cite{elmore2017pathologists}).

A tool that allows labs to sort and prioritize melanoma cases in advance of pathologist review could improve turnaround time, enabling pathologists to review cases requiring faster turnaround time early in the day. This is particularly important as shorter turnaround time is correlated with improved overall survival for melanoma patients \cite{matthewsmel}.  It could also alleviate common lab bottlenecks such as referring cases to specialized dermatopathologists, or ordering additional tissue staining beyond the standard H\&E. These contributions are especially important as the number of skin biopsies performed per year has skyrocketed, while the number of practicing pathologists has declined \cite{metter2019trends}.

The advent of digital pathology has brought the revolution in machine learning and artificial intelligence to bear on a variety of tasks common to pathology labs. Several deep learning algorithms have been introduced to distinguish between different skin cancers and healthy tissue with very high accuracy (e.g. \cite{de2020recognition, thomas2021interpretable, zormpas2021superhistopath,litjensdl}). However, almost all of these studies fail to demonstrate the robustness required for use in a clinical workflow setting  because they were tested a on small number ($<\sim$1000) of WSIs. Moreover, these algorithms are often not capable of triaging WSIs, as they use curated training and test datasets that do not represent the diversity of cases encountered in a dermatopathology lab. Many of them rely on pixel-level annotations to train their models, which is slow and expensive to scale to a large dataset with greater variability. 
Considerable advancements have been made towards systems capable of use in clinical practice for prostate cancer. Campanella \textit{et al.} \cite{campanella2019clinical} trained a model in a weakly-supervised framework that did not require pixel-level annotations to classify prostate cancer and validated on $\sim10,000$ WSIs sourced from multiple countries.  However, some degree of human-in-the-loop curation was performed on their dataset, including manual quality control such as post-hoc removal of slides with pen ink from the study. Pantanowitz \textit{et al.} \cite{ibex} used pixel-wise annotations to develop a model trained on $\sim550$ WSIs that distinguishes high-grade from low-grade prostate cancer. In dermatopathology, the model developed in \cite{ianni2020tailored} classified skin lesion specimens between four morphology-based groups, was tested on $\sim13,500$ WSIs, and also demonstrated that use of confidence thresholding could provide a high accuracy; however, it grouped malignant melanoma with all other benign melanocytic lesions, limiting its potential uses. Additionally, all previous attempts at pathology classification using deep learning have, at their greatest level of abstraction, performed classification at the level of a WSI or a sub-region of a WSI. Since a pathologist is required to review all WSIs from a tissue specimen, previous deep learning pathology efforts therefore do not leverage the same visual information that a pathologist would have at hand to perform a diagnosis, require some curation of datasets to ensure that pathology is present in all training slides, and implement \textit{ad-hoc rules} for combining the predictions of each WSI corresponding to a specimen. Most have also neglected the effect of diagnostic discordance on their ground truth, resulting in potentially mislabeled training and testing data.

In this work, we present a pathology deep learning system (PDLS) that can classify skin cases for triage and prioritization prior to pathologist review. Unlike previous systems, our PDLS performs hierarchical melanocytic specimen classification into low (MPATH I-II), Intermediate (MPATH III), or High (MPATH IV-V) diagnostic categories, enabling prioritization of melanoma cases. Our PDLS is the first to classify skin biopsies at the specimen level through a collection of WSIs that represent the entirety of the tissue from a single specimen. This training procedure ultimately models the process of a dermatopathologist, who reviews the full collection of scanned WSIs corresponding to a specimen to make a diagnosis.  Finally, we train and validate the PDLS on the largest dataset of consensus-reviewed melanocytic specimens published to date. The system is built to be scalable and ready for the real-world, built without any pixel-level annotations, and incorporating the automatic removal of scanning artifacts.

\begin{table}[]
\centering
\begin{tabular}{lr}
\toprule
                             Diagnostic Morphology &  Counts \\
\midrule
                       \textbf{Basaloid} & 544 \\
                       \hline
		      Nodular Basal Cell Carcinoma &     404 \\
		      Basal Cell Carcinoma, NOS & 123 \\
                Basal Cell Carcinoma, Morphea type &       7 \\
                                     Pilomatrixoma &       5 \\
                Infiltrative Basal Cell Carcinoma &       5 \\
		 \hline
		 \textbf{Squamous} & 530 \\
		 \hline
                  Invasive Squamous Cell Carcinoma &     269 \\
                  Squamous Cell Carcinoma in situ &     254 \\
                   (Bowen's Disease) \\                  Fibrokeratoma &       4 \\
                                Warty Dyskeratorma &       3 \\
					\hline
				 \textbf{Melanocytic High Risk} & 102 \\
 					\hline
                                          Melanoma &      102 \\
				\hline
				\textbf{Melanocytic Intermediate Risk} & 213\\
				\hline
                                  Melanoma In Situ &     202 \\
                                  Severe Dysplasia &       9 \\
                                   \hline
				  \textbf{Melanocytic Low Risk} & 764\\
				  \hline
				   Conventional Melanocytic Nevus & 368 \\
				    (acquired and congenital)\\
				     Mild Dysplasia &     289 \\
		   Moderate Dysplasia &      75 \\ 
            
                    Halo Nevus & 14 \\
                             Dysplastic Nevus, NOS &      12 \\
   Spitz Nevus & 2 \\
                                        Blue Nevus &       2 \\
				\hline
				\textbf{Other Diagnoses} & 1360 \\ 
\bottomrule
\end{tabular}
 \caption{Counts of each of the general pathologies in the reference set, broken-out into specific diagnostic entities.}
\label{tbl:subtype_labels_UFCM}
 \end{table}

\section{Methods}
\label{sec:methods}

\subsection{Reference and Validation Lab Data Collection}
\label{subsec:data_collection}
%
The PDLS was trained using slides from 3511 specimens (consisting of 7685 WSIs) collected from a leading dermatopathology lab in a top academic medical center (Department of Dermatology at University of Florida College of Medicine),  which we refer to as the \textit{`Reference Lab'}. The Reference Lab dataset consisted of both an uninterrupted series of sequentially-accessioned cases (69\% of total specimens) and a targeted set, curated to enrich for rarer melanocytic pathologies (31\% of total specimens). 
Melanocytic specimens were only included in this set if 3 dermatopathologists' consensus on diagnosis could be established. 
The WSIs consisted exclusively of H\&E-stained, formalin-fixed, paraffin-embedded dermatopathology tissue and were scanned using a 3DHistech P250 High Capacity Slide Scanner at an objective power of 20X, corresponding to 0.24$\mu$m/pixel. The final classification given by the PDLS was one of six classes, defined by their morphologic characteristics --
\begin{enumerate} 
\item Basaloid: containing abnormal proliferations of basaloid-oval cells, primarily basal cell carcinoma of various types
\item Squamous: containing malignant squamoid epithelial proliferations, consisting primarily of squamous cell carcinoma (invasive and in situ)
\item Melanocytic Low Risk: benign to moderately atypical melanocytic nevi/proliferation of cells of melanocytic origin, classified as the MPATH I or MPATH II diagnostic category
\item Melanocytic Intermediate Risk: severely atypical melanocytic nevi or melanoma in situ, classified as the MPATH III diagnostic category
\item Melanocytic High Risk: invasive melanoma, classified as the MPATH IV or V diagnostic category
\item Other: all skin specimens that do not fit into the above classes, including but not limited to inflammatory conditions and benign proliferations of squamoid epithelial cells.
\end{enumerate}
The overall reference set was composed of 544 Basaloid, 530 Squamous, 1079 Melanocytic and 1358 Other specimens. Of the Melanocytic specimens, 764 were Low Risk , 213 were Intermediate Risk and 102 were High Risk. The heterogeneity of this reference set is illustrated in Table~\ref{tbl:subtype_labels_UFCM}.  The specimen counts that we report for the melanocytic classes throughout this work reflect counts following three-way consensus review (see Section~\ref{subsec:discordance_study}). For training, validating, and testing the PDLS, we divided this dataset into three partitions by sampling at random without replacement with 70\% of specimens used for training, and 15\% used for each of validation and testing.

%
%
To validate model performance and generalizability across labs, scanners, and associated histopathology protocols, we collected several large datasets of similar composition to the Reference Lab from leading dermatopathology labs of two additional top academic medical centers (Jefferson Dermatopathology Center, Department of Dermatology \& Cutaneous Biology, Thomas Jefferson University, denoted as \textit{Validation Lab 1}, and Department of Pathology and Laboratory Medicine at Cedars-Sinai Medical Center, which is denoted as \textit{Validation Lab 2}). These datasets both comprised of: 1) an uninterrupted set of sequentially-accessioned cases - 65\% for Validation Lab 1, 24\% for Validation Lab 2) a set targeted to heavily sample melanoma, pathologic entities that mimic melanoma, and other rare melanocytic specimens. Specimens from \textit{Validation Lab 1} consisted of slides from 2795 specimens (3033 WSIs), scanned using a 3DHistech P250 High Capacity Slide Scanner at an objective power of 20X (0.24 $\mu$m/pixel). Specimens from \textit{Validation Lab 2} consisted of slides from 2066 specimens (2066 WSIs; each specimen represented by a single WSI), with WSIs scanned using a Ventana DP 200 scanner at an objective power of 20X (0.47 $\mu$m/pixel). Note: specimen and WSI counts above reflect specimens included in the study \textit{after} screening melanocytic specimens for inter-pathologist consensus (see \ref{subsec:discordance_study}). Table~\ref{tab:labels_per_lab} shows the class distribution for the Validation labs.

\begin{table}[]

    \centering
    \begin{tabular}{l|cc}
    Label category & Validation Lab 1 & Validation Lab 2\\
    \hline
    MPATH I-II     &  1457 & 458 \\
    MPATH III     &  225 & 364 \\
    MPATH IV-V     & 100 & 361 \\
    Basaloid     & 198 & 265 \\
    Squamous     &  104 & 55 \\
    Other    & 711 & 563 \\
    \end{tabular}
    \caption{Class counts for the Validation Lab datasets. (The counts for the Reference Lab are given in Table \ref{tbl:subtype_labels_UFCM}.)}
    \label{tab:labels_per_lab}
\end{table}


\subsection{Consensus Review}
\label{subsec:discordance_study}
There are high discordance rates in diagnosing melanocytic specimens. Elmore \textit{et al.} \cite{elmore2017pathologists} studied 240 dermatopathology cases and found that the consensus rate for MPATH Class II lesions was 25\%, for MPATH Class III lesions 40\%, and for MPATH Class IV 45\% . Therefore, three board-certified pathologists reviewed each melanocytic specimen to establish a reliable ground truth for melanocytic cases in our study. The first review was the original specimen diagnosis made via glass slide examination under a microscope. Two additional dermatopathologists independently reviewed and rendered a diagnosis digitally for each melanocytic specimen. The patient's year of birth and gender were provided with each specimen upon review. Melanocytic specimens were considered to have a consensus diagnosis and included in the study if
\begin{enumerate}
    \item all three dermatopathologists were in consensus on a diagnostic class for the specimen.
    \item two of three dermatopathologists were in consensus on a diagnostic class for the specimen \textit{and} a fourth and fifth pathologist reviewed the specimen digitally and both agreed with the majority classification.
\end{enumerate} 
A diagnosis was rendered in the above fashion for every melanocytic specimen obtained from the Reference Lab and Validation Lab 1. All dysplastic and malignant melanocytic specimens from Validation Lab 2 were reviewed by three dermatopathologists and only the specimens for which consensus could be established were included in the study. No non-melanocytic specimens were reviewed for concordance due to inherently lower known rates of discordance \cite{bccdiscordance}.

For the specimens obtained from the Reference Lab, consensus was established for 75\% of specimens originally diagnosed as MPATH I/II, 66\% of those diagnosed as MPATH III, 87\% of those diagnosed as MPATH IV/V, and for 74\% of the reviewed specimens in total. For specimens obtained from Validation Lab 1, pathologists consensus was established for 84\% of specimens originally diagnosed as MPATH I/II specimens, 51\% of those diagnosed as MPATH III, 54\% of those diagnosed as MPATH IV/V, and for 61\% of the reviewed specimens in total.


\subsection{Pathology Deep Learning System Architecture}
\label{subsec:model_architecture}

\begin{figure*}[tb]
    \centering
    \includegraphics[width=\textwidth]{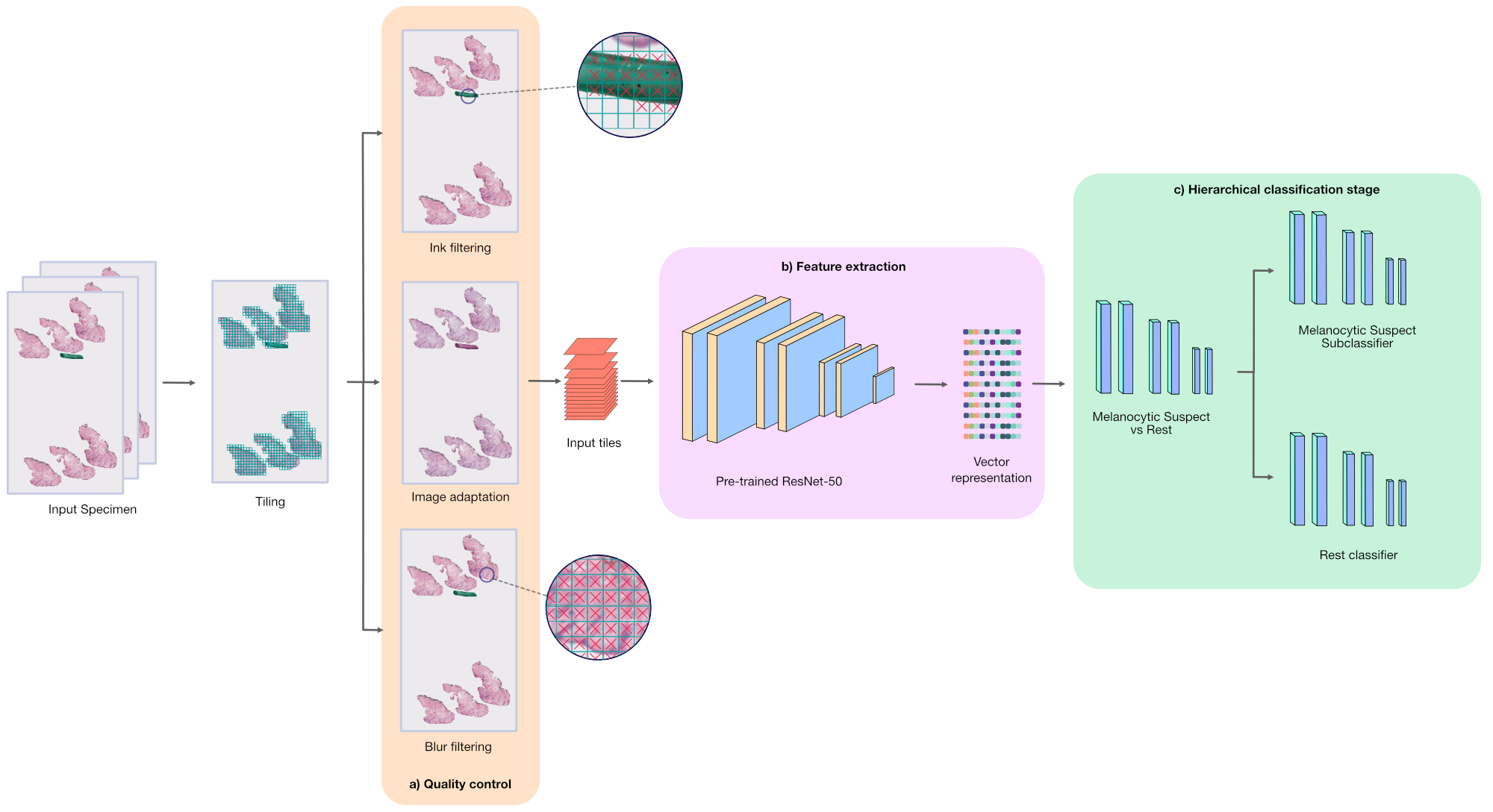}
    \caption{The stages of the PDLS are: Quality Control, Feature Extraction, and Hierarchical Classification. All single specimen WSIs were first passed through the tiling stage, then the quality control stage consisting of ink filtering, blur filtering and image adaptation. The image-adapted tiles were passed through the feature extraction stage consisting of a pretrained ResNet50 network to obtain embedded vectors. Finally, the vectors were propagated through the hierarchical classification stage consisting of an upstream model performing a binary classification between ``Melanocytic Suspect'' and ``\textit{Rest}''. Specimens that were classified as ``Melanocytic Suspect" were fed into a downstream model, which classified between ``Melanocytic High Risk, Melanocytic Intermediate Risk" and ``\textit{Rest}".  The remaining specimens were fed into a separate downstream ``\textit{Rest}'' model, which classified between ``Basaloid, Squamous, Melanocytic Low Risk" and ``Other".}
    \label{fig:PDLS_pipeline}
\end{figure*}

The PDLS consists of three main components: quality control, feature extraction and hierarchical classification. A diagram of the PDLS is shown in Figure \ref{fig:PDLS_pipeline}.  Each specimen was first segmented into tissue-containing regions, subdivided into 128x128 pixel tiles, and extracted at an objective power of 10X. Each tile was passed through the quality control and embedding components of the PDLS.

Quality control consisted of ink filtering, blur filtering, and image adaptation. Pen ink is common in labs migrating their workload from glass slides to WSIs where the location of possible malignancy was marked. This pen ink represented a biased distractor signal in training the PDLS that is highly correlated with malignant or High Risk pathologies. Tiles containing pen ink were identified by a weakly supervised model trained to detect inked slides. These tiles were removed from the training and validation data and before inference on the test set. We also sought to remove areas of the image that were out of focus due to scanning errors by setting a threshold on the variance of the Laplacian over each tile \cite{pertuz2013AnalysisOF, pech2000diatom}. In order to avoid domain shift between the colors of the training data and validation data \cite{ciompi2017importance, stacke2019closer, tellez2019quantifying}, we adopted the image adaptation procedure in \cite{ianni2020tailored}.


The next component of the PDLS extracted informative features from the quality controlled, color-standardized tiles. To capture higher-level features in these tiles, they were propagated through a neural network (ResNet50; \cite{resnet50}) trained on the ImageNet \cite{imagenet} dataset to embed each input tile into 1024 channel vectors which were then used in our subsequent models.


We developed a hierarchical model architecture in order to classify both Melanocytic High and Intermediate Risk specimens with high sensitivity. First, an upstream model performed a binary classification between ``Melanocytic Suspect" (defined as ``High or Intermediate Risk")  and ``Basaloid, Squamous, Low Risk" or ``Other" (which we collectively define as the ``\textit{Rest}'' class). Specimens that were classified as ``Melanocytic Suspect" were fed into a downstream model, which further classified the specimen between ``Melanocytic High Risk, Melanocytic Intermediate Risk" and ``\textit{Rest}". The remaining specimens, classified as ``\textit{Rest}'', were fed into a separate downstream model, which further classified the specimen between ``Basaloid, Squamous, Melanocytic Low Risk" and ``Other". Each model consisted of four fully-connected layers (two layers of 1024 channels each, followed by two of 512 channels each). Each neuron in the three layers after the input layer was ReLU activated. The three models in the hierarchy were trained under a weakly-supervised multiple-instance learning (MIL) paradigm. Each embedded tile was treated as an instance of a bag containing all quality-assured tiles of a specimen. Embedded tiles were aggregated using sigmoid-activated attention heads \cite{ilse2018attention, lu2020clam}. To help prevent over-fitting, the training dataset consisted of augmented versions (inspired by Tellez \textit{et al.}\cite{tellez}) of the tiles. Augmentations were generated with the following augmentation strategies: random variations in brightness, hue, contrast, saturation, (up to a maximum of 15\%),  Gaussian noise with 0.001 variance, and random 90 degree image rotations.  The upstream binary ``Melanocytic Suspect vs. \textit{Rest}'' classification model and the downstream ``\textit{Rest}" subclassifier were each trained end-to-end with cross-entropy loss. The ``Melanocytic Suspect'' subclassifier was also trained with cross-entropy loss, but with a multi-task learning strategy. This subclassifier was presented with three tasks: differentiating ``Melanocytic High Risk" from ``Melanocytic Intermediate Risk" specimens,  ``Melanocytic High Risk'' from ``\textit{Rest}'' specimens, and  ``Melanocytic Intermediate Risk'' from ``\textit{Rest}'' specimens. The training loss for this subclassifier was computed for each task, but was masked if it did not relate to the ground truth label of the specimen. Two out of three tasks were trained for any given specimen in a training batch. By training in this manner, we were able to use the shared network layers as a generic representation of melanocytic pathologies, while the task branches learned to attend to specific differences to accomplish their tasks.

\begin{figure}
    \centering
    \includegraphics[width=0.5\textwidth]{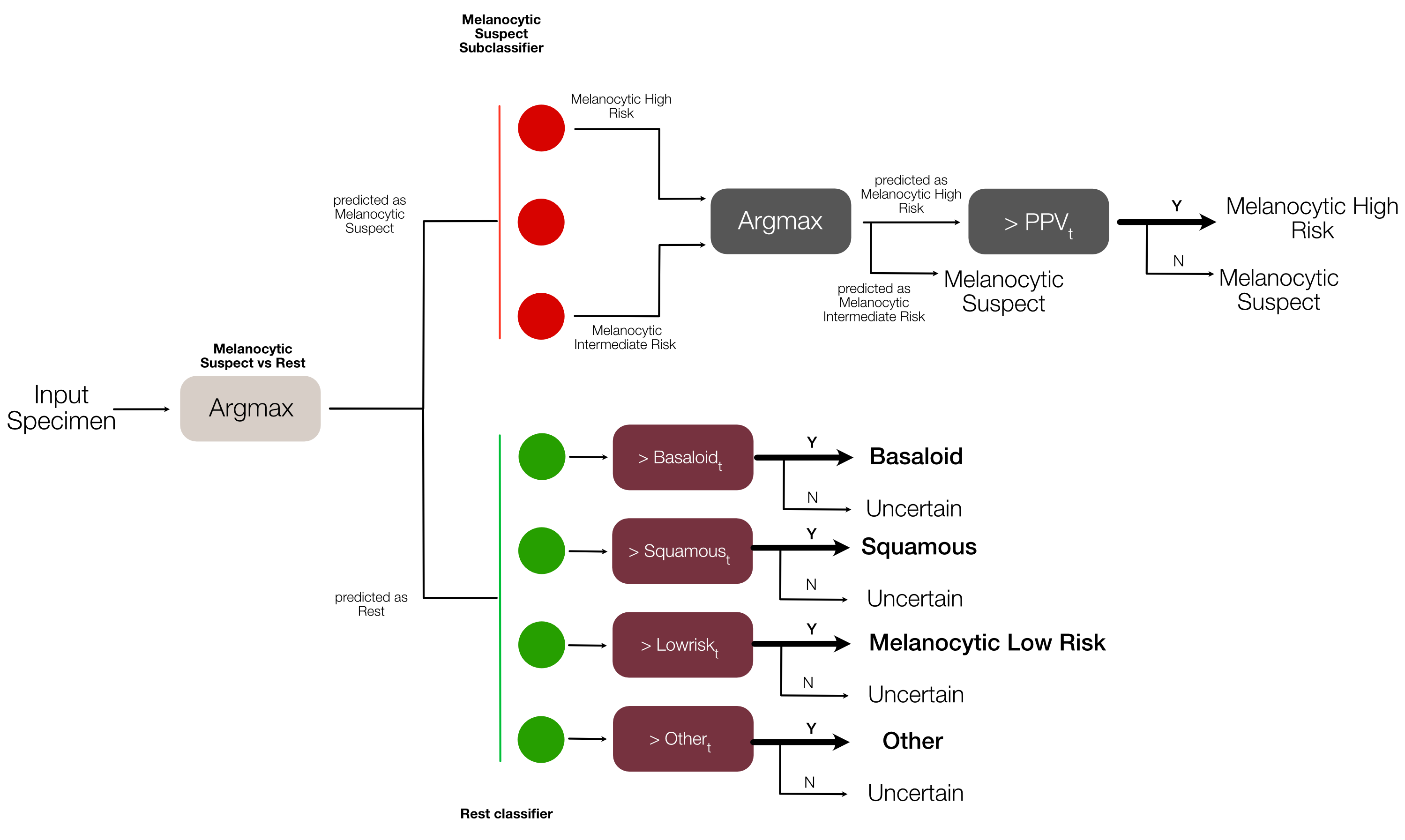}
    \caption{The hierarchical model prediction procedure for a given input specimen. Specimens that were predicted to be ``Melanocytic High Risk'' had to pass both an accuracy threshold and a Positive Predictive Value (PPV) threshold (``PPV$_{t}$'') --- both established a priori on the validation set-- to be predicted as ``Melanocytic High Risk''; otherwise they were predicted as ``Melanocytic Suspect''. The confidence thresholding procedure is described in the text.}
    \label{fig:prediction_flow}
\end{figure}

During inference, an input specimen's predicted classes were computed as follows (see Figure \ref{fig:prediction_flow}):
\begin{enumerate}
    \item The larger of the two confidence values (see below for our confidence thresholding procedure) output from the upstream classifier determined which downstream classifier a specimen was handed to.
    \item If the specimen was handed to the ``\textit{Rest}'' subclassifier, we used the highest confidence class probability as the predicted label.
    \item If the specimen was handed to the Melanocytic Suspect subclassifier, we used the highest confidence class probability between the ``Melanocytic High Risk vs. \textit{Rest}'' and ``'Melanocytic Intermediate Risk vs \textit{Rest}'' tasks as the predicted label.
\end{enumerate}

As a final step in the classification pipeline, we performed classification with uncertainty quantification to establish a confidence score for each prediction using a Monte Carlo dropout method following the same procedure as used by Gal \& Ghahramani \cite{gal2016dropout}. Using the confidence distribution of the specimens in the validation set of the Reference Lab, we computed confidence threshold values for each predicted class following the procedure outlined Ianni et al.\cite{ianni2020tailored} by requiring classifications to meet a predefined a level of accuracy in the validation set.  Specimens that were predicted as ``Melanocytic High Risk'' had to pass two confidence thresholds: an accuracy threshold and a PPV threshold in order to be predicted as ``Melanocytic High Risk''.  Specimens that were predicted to be ``Melanocytic High Risk`` but failed to meet these thresholds were predicted as ``Melanocytic Suspect``. We set thresholds that maximized the sensitivity of the PDLS to the ``Melanocytic Suspect'' class, while simultaneously maximizing the PPV to the``Melanocytic High Risk'' class.   

To evaluate how our PDLS generalizes to data from other labs, we fine-tuned the model trained on data from the Reference Lab to both Validation Lab 1 and Validation Lab 2. We specifically set aside 255 specimens from each validation lab (using an equal class distribution of specimens) as our \textit{calibration} set, of which we used 210 specimens as the training set and 45 specimens as the validation set for fine-tuning the models. (The remaining  specimens in the validation lab used as our test set.) Our final validation lab metrics in Section \ref{subsec:performance} are reported on the test set with these calibrated models. 


\section{Results}
\label{sec:results}

\subsection{PDLS Performance Evaluation}
\label{subsec:performance}
\begin{figure*}[tb]
    \centering
    \includegraphics[width=0.6\columnwidth]{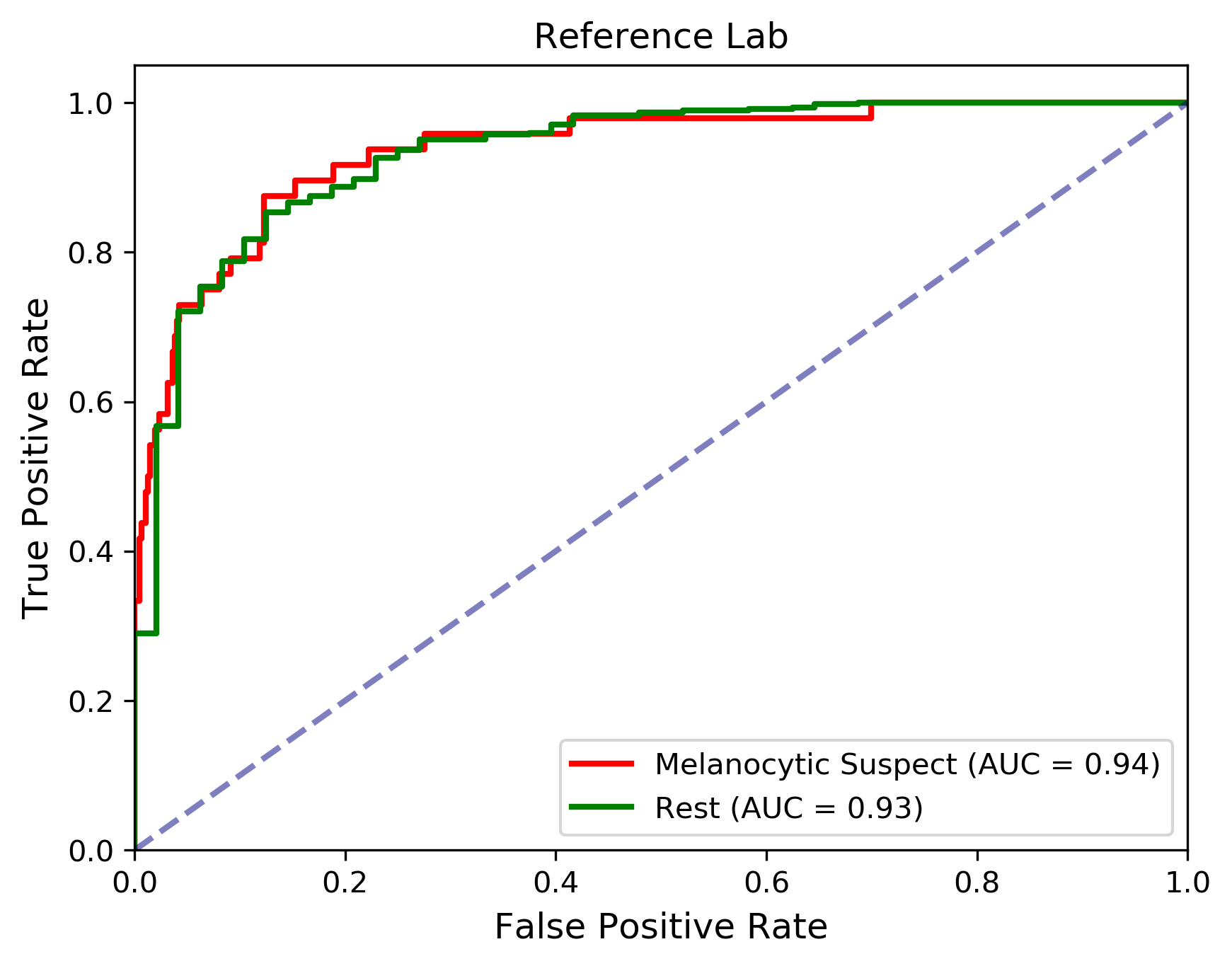}
    \includegraphics[width=0.6\columnwidth]{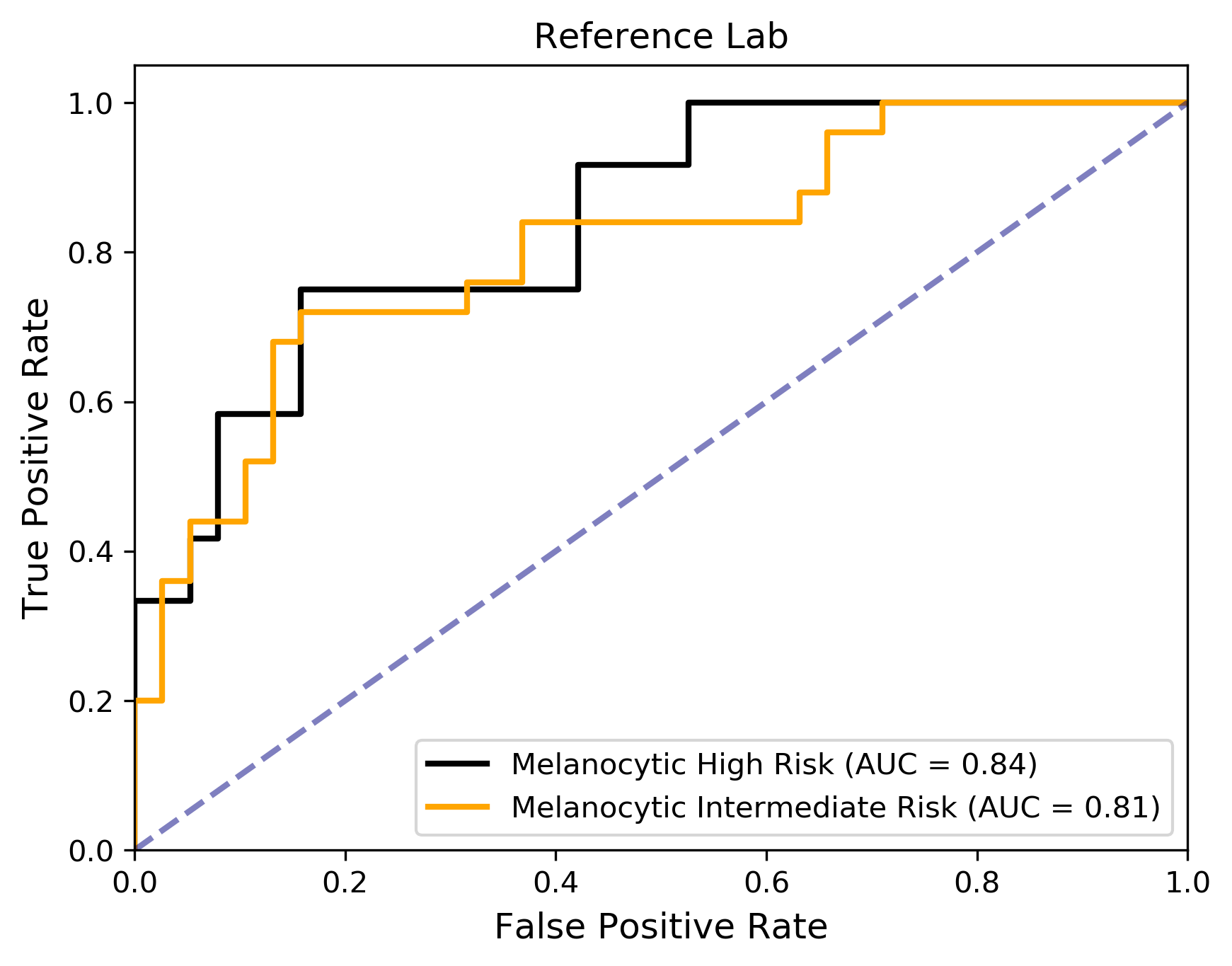}
    \includegraphics[width=0.6\columnwidth]{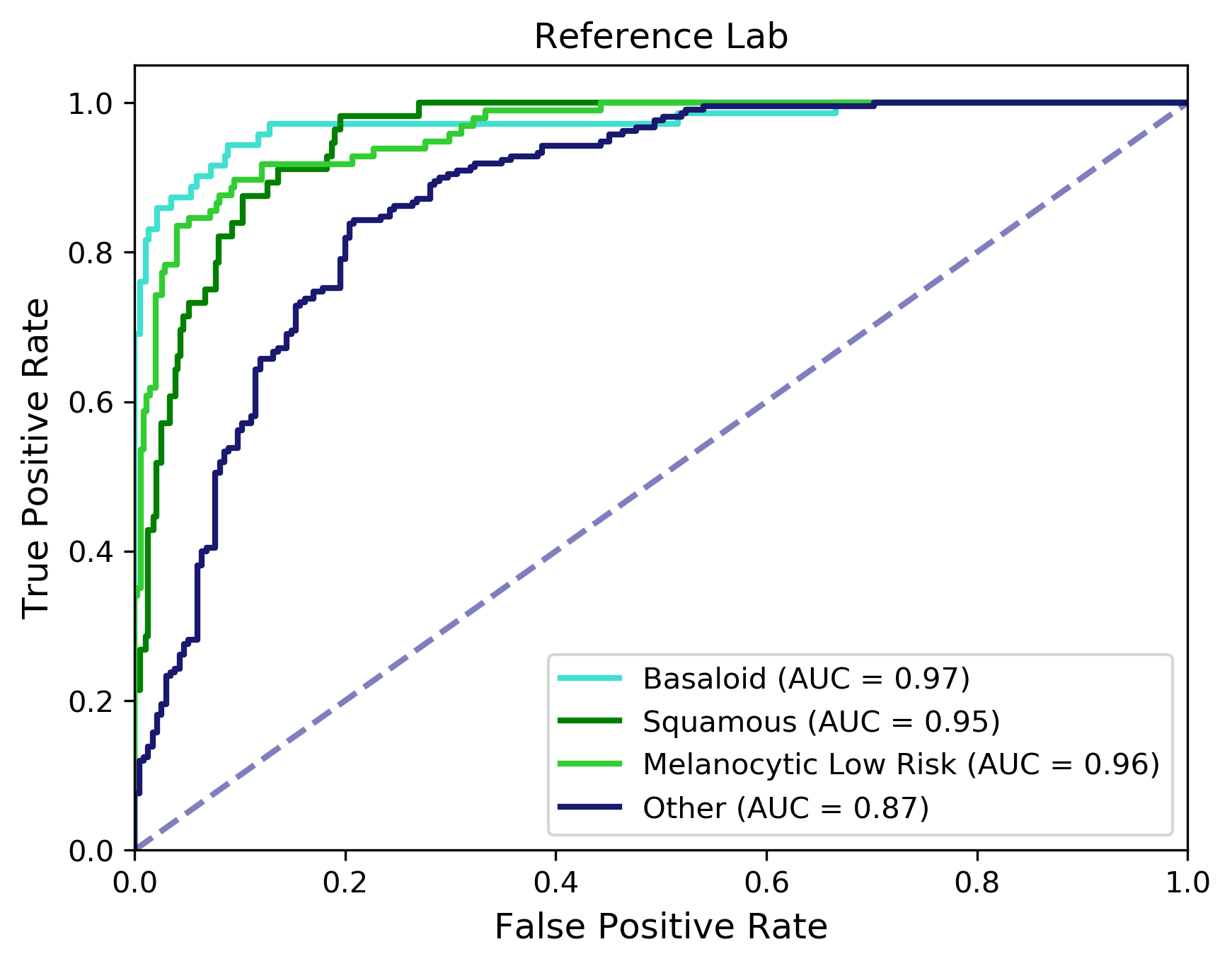}
    \includegraphics[width=0.6\columnwidth]{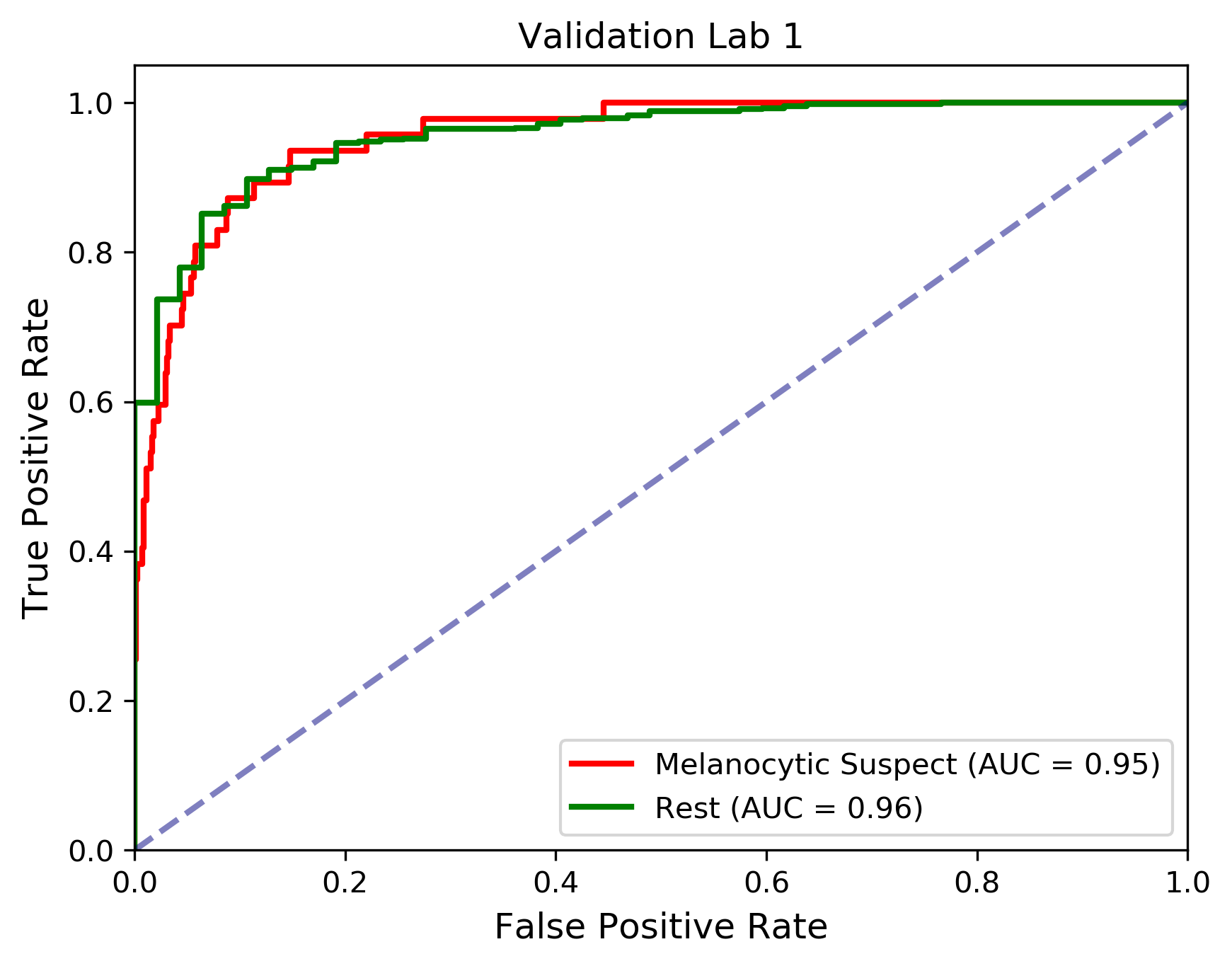}
    \includegraphics[width=0.6\columnwidth]{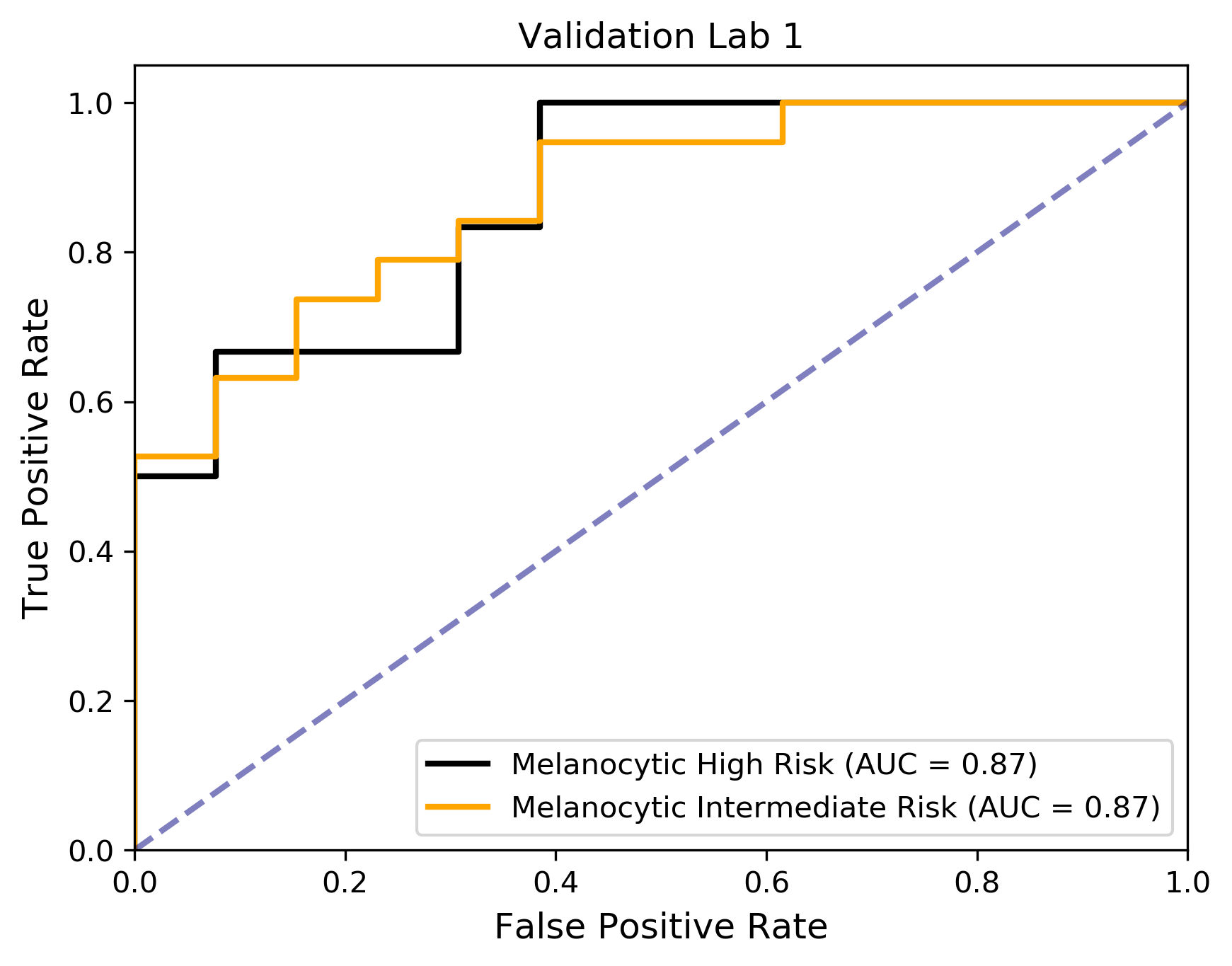}
    \includegraphics[width=0.6\columnwidth]{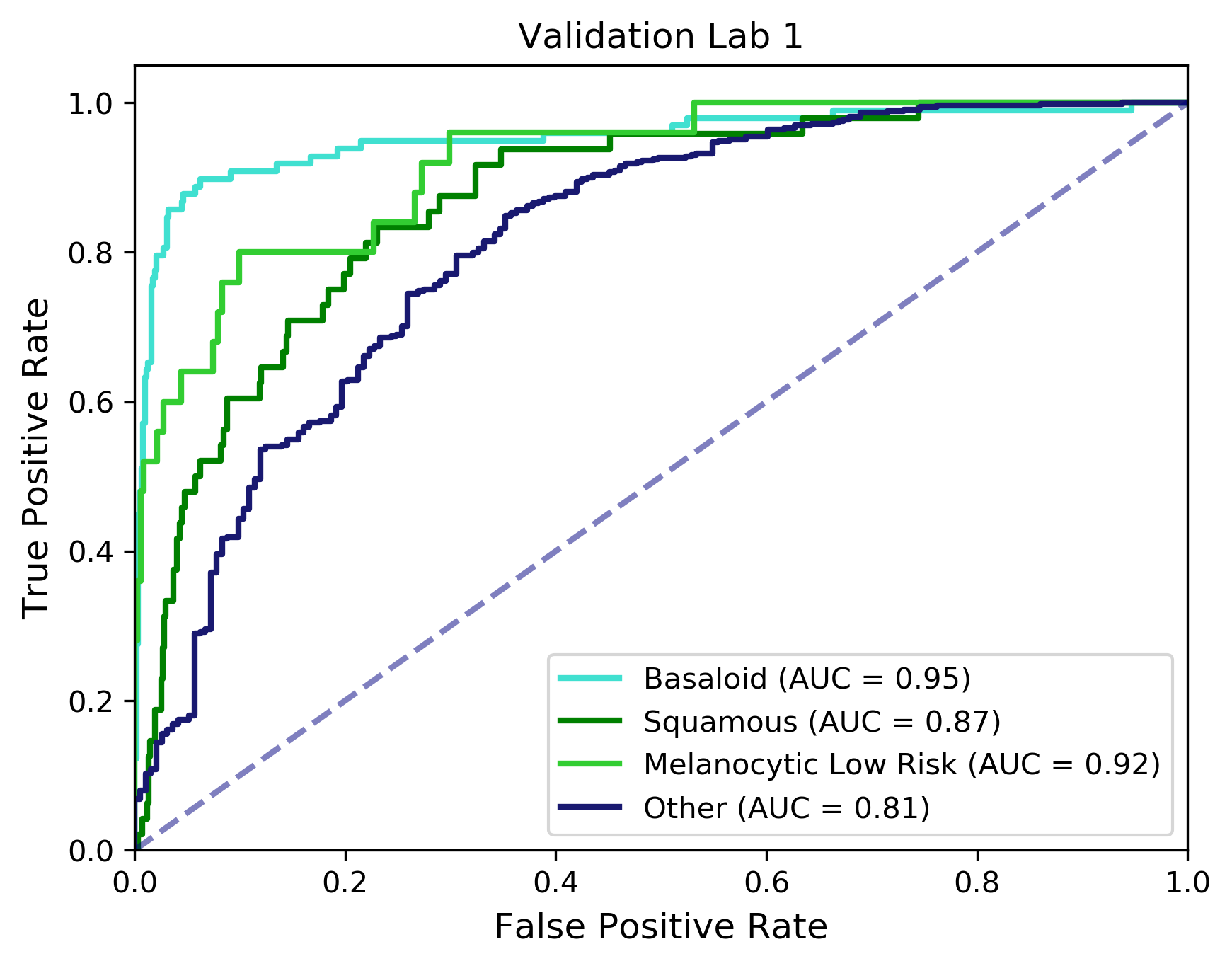}
    \includegraphics[width=0.6\columnwidth]{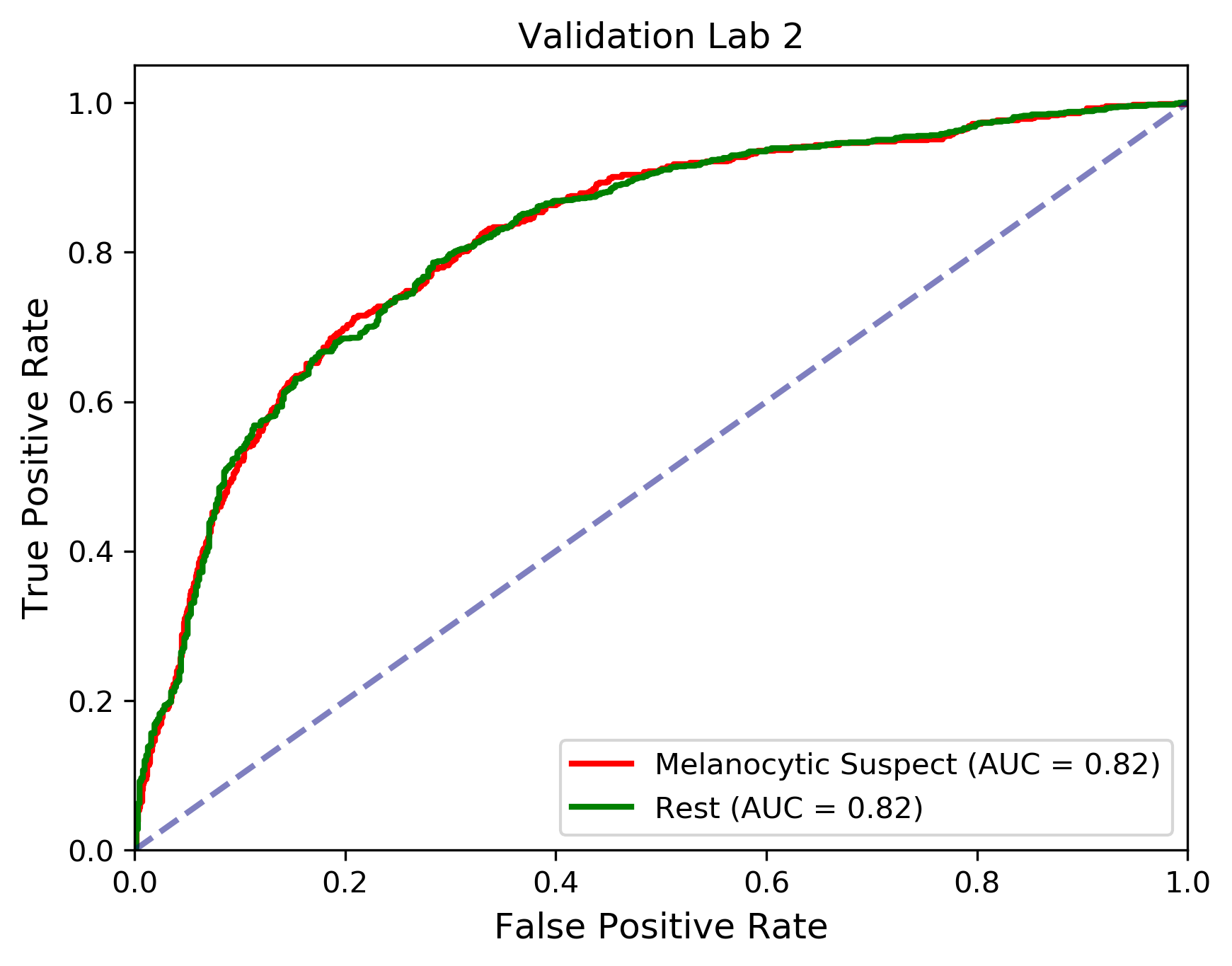}
    \includegraphics[width=0.6\columnwidth]{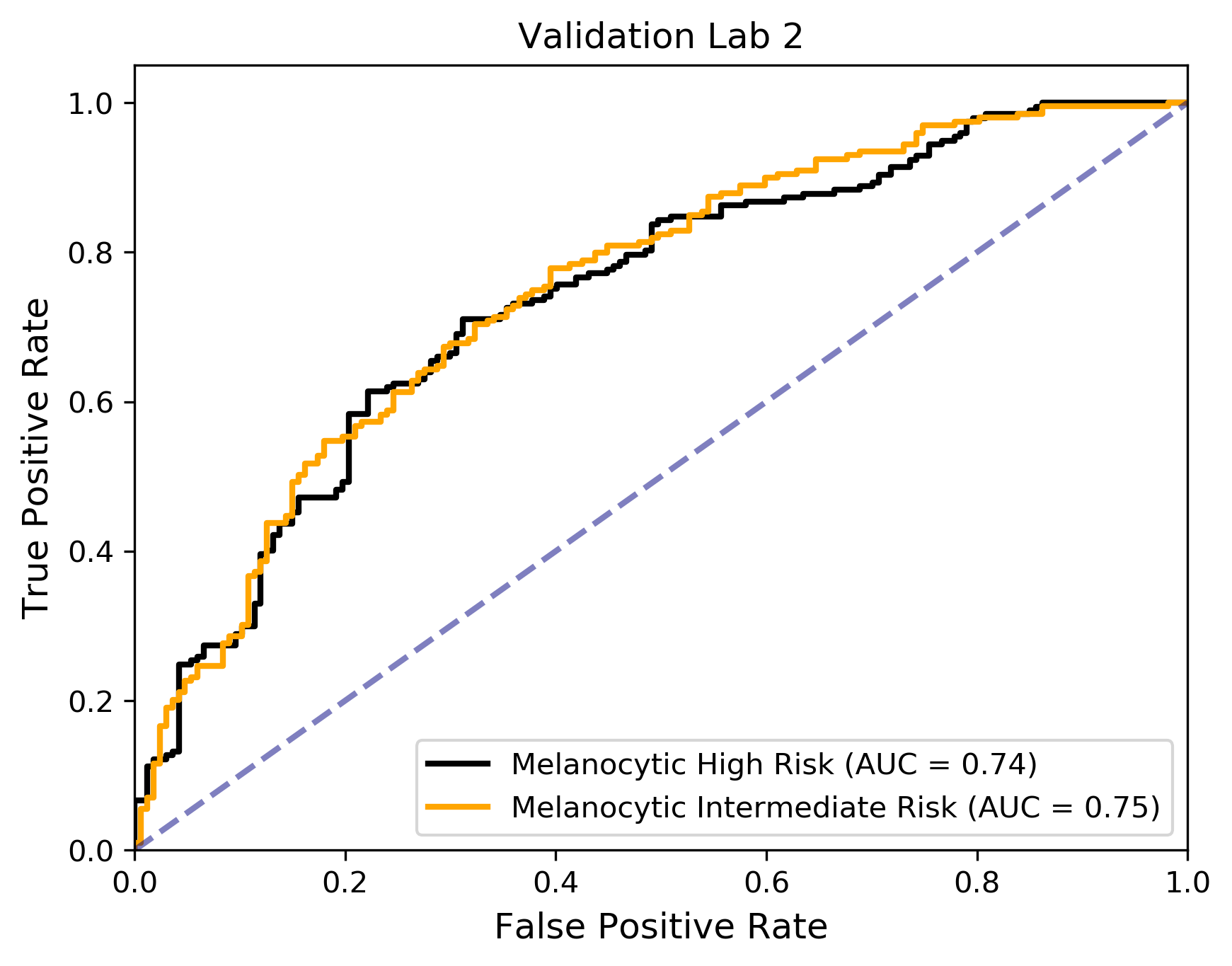}
    \includegraphics[width=0.6\columnwidth]{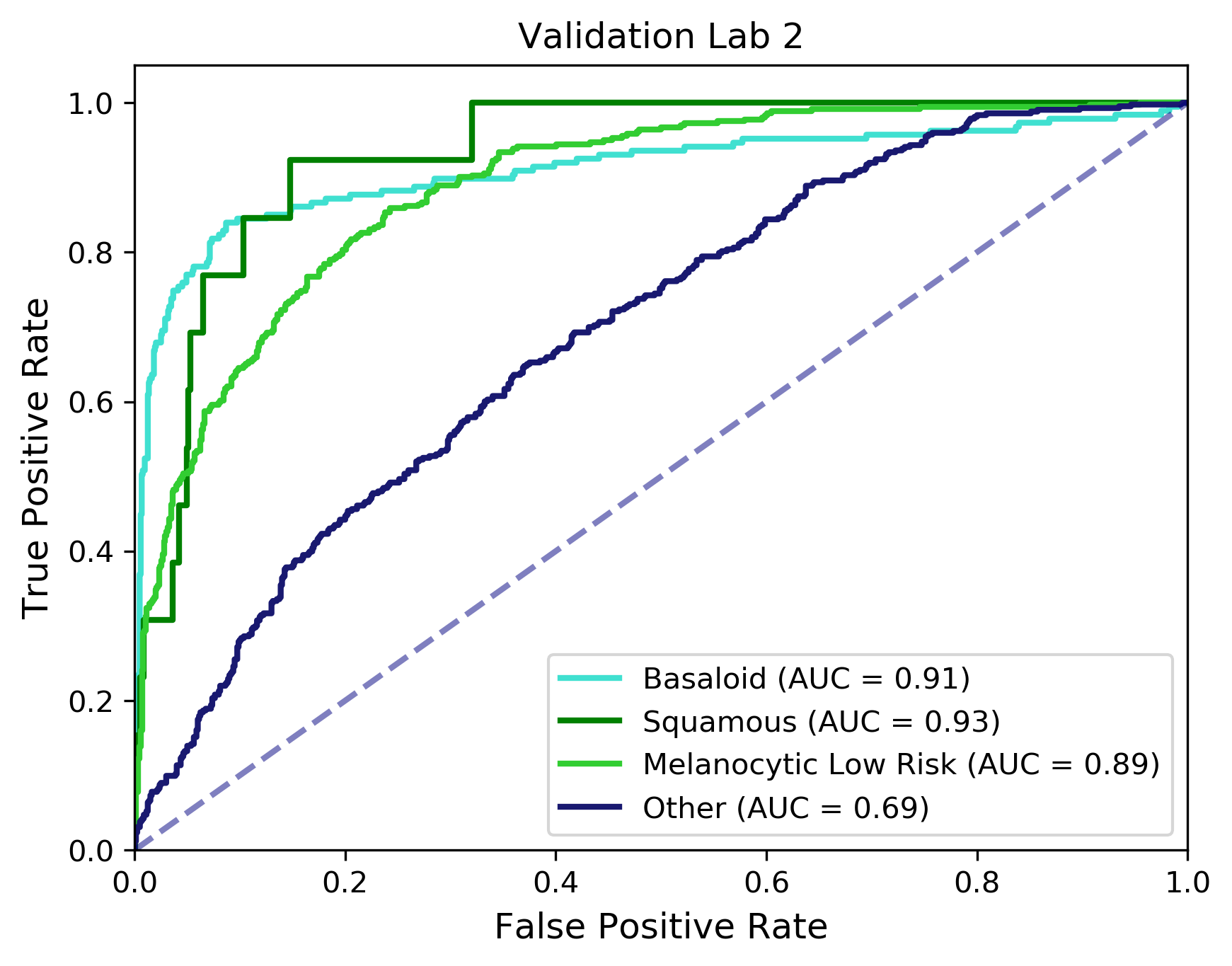}
    \caption{Receiver Operating Characteristic (ROC) curves for the individual models - the upstream classifier (\textit{left column}), the High \& Melanocytic Intermediate classifier (\textit{middle column}) and the Basaloid, Squamous, Low Risk Melanocytic \& \textit{Rest} classifier (\textit{right column}) - from the hierarchical architecture of the PDLS are shown for the Reference Lab (\textit{first row}), the first Validation Lab, (\textit{second row}) and the second Validation Lab (\textit{third row}).}
    \label{fig:results}
\end{figure*}

To demonstrate the performance of our PDLS, we first show Receiver Operating Characteristic (ROC) curves for the individual hierarchical component models. The ROC curves derived from the Reference Lab test dataset are shown in Figure \ref{fig:results}. The Area Underneath the ROC Curve (AUC) values, calculated with the one-vs-rest scoring scheme, were \textbf{0.97, 0.95, 0.87, 0.84, 0.81, 0.93, and 0.96} for the Basaloid, Squamous, Other, Melanocytic High Risk, Melanocytic Intermediate Risk, Melanocytic Suspect, and Melanocytic Low Risk classes, respectively.  Table~\ref{tab:specific_label_performance} shows the PDLS performance with respect to diagnostic entities of clinical interest on the Reference lab test dataset. The sensitivity of the PDLS to the Melanocytic Suspect class was found to be \textbf{0.83, 0.85} for the  Melanocytic High and Intermediate risk classes, respectively. The PPV to Melanocytic High Risk was found to be 0.57.  The dropout Monte Carlo procedure set the threshold for Melanocytic High Risk classification very high; specimens below this threshold were classified as Melanocytic Suspect, maximizing the sensitivity to this class.   

After fine-tuning all three models in the hierarchy through the \textit{calibration} procedure in each validation lab, we report the ability of our PDLS to generalize to unseen data from both validation labs. Note that fine-tuning was not performed for any of the models in the pre-processing pipeline (Colorization, Ink Detection or ResNet). The ROC curves derived from the Validation Lab 1 and Validation Lab 2 test datasets are shown in Figure \ref{fig:results}. The AUC values for Validation Lab 1 were \textbf{0.95, 0.88, 0.81,0.87, 0.87, 0.95, and 0.92} for the Basaloid, Squamous, Other, Melanocytic High Risk, Intermediate Risk, Suspect, and Low Risk classes, respectively and the AUC values for the same classes for Validation Lab 2 were \textbf{0.93, 0.92, 0.69, 0.76, 0.75, 0.82, and 0.92}.  

\begin{table*}[]
    \centering
    \begin{tabular}{lrrrrr}
\toprule
Diagnosis &  PPV &  Sensitivity &  F$_1$ Score &  Balanced Accuracy & Support\\
\midrule
Melanoma $\rightarrow$ Melanocytic High Risk       &       0.66 &    0.45 &    0.47 &    0.52 & 23\\
Melanoma $\rightarrow$ Melanocytic Suspect & 1.00 & 0.83 & 0.90 & 0.83 & 23 \\
Melanoma in situ $\rightarrow$ Melanocytic Intermediate Risk &       1.00 &    0.75 &    0.86 &    0.75 & 20\\
Melanoma in situ $\rightarrow$ Melanocytic Suspect &       1.00 &    0.85 &    0.92 &    0.85 & 20\\
Spitz  Nevus     &       0.00 &    0.00 &    0.00 &    0.00 &      2 \\
Dysplastic Nevus  &       0.91 &    0.76 &    0.82 &    0.56 & 61\\
Dermal Nevus      &       1.00 &    0.81 &    0.90 &    0.81 & 28 \\
Compound Nevus    &       0.94 &    0.75 &    0.82 &    0.55 & 73\\
Junctional Nevus  &       0.84 &    0.77 &    0.80 &    0.42 & 61\\
Halo Nevus             &       1.00 &    1.00 &    1.00 &    1.00 & 20\\
Blue Nevus             &       1.00 &    0.67 &    0.80 &    0.67 & 68\\
Squamous Cell Carcinoma  &       1.00 &    0.81 &    0.89 &    0.81 & 15\\
Bowen's Disease &       1.00 &    0.85 &    0.92 &    0.85 & 4\\
Basal Cell Carcinoma &       1.00 &    0.84 &    0.91 &    0.84 & 8\\
\bottomrule
\end{tabular}
    \caption{Metrics for selected diagnoses of clinical interest, based on the Reference Lab test set, representing the classification performance of the individual diagnoses into their higher-level classes: e.g., a correct classification of ``Melanoma" is the prediction ``Melanocytic High Risk".  Results are class-weighted according to the relative prevalence in the test set.}
    \label{tab:specific_label_performance}
\end{table*}

\subsection{Consensus Ablation Study}
\begin{figure}
    \centering
    \includegraphics[scale=0.40]{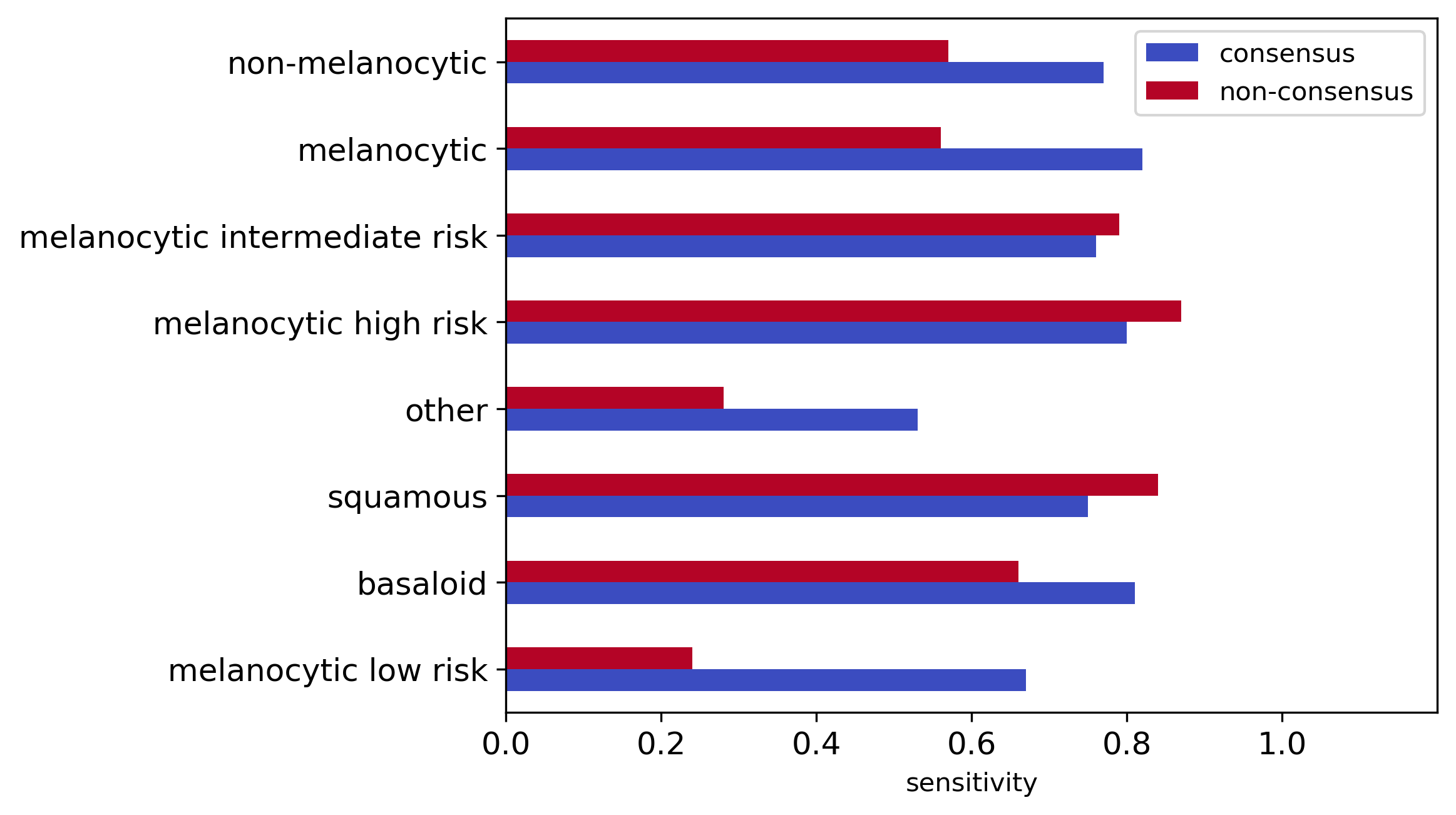}
    \caption{Reference lab performance on the same test set when trained on consensus and non-consensus data. The melanocytic class is defined as the Low, Intermediate and High Risk classes. The sensitivity of the Melanocytic Intermediate and High Risk classes are defined with respect to the PDLS classifying these classes as suspect. The PPV to melanocytic high risk in the non-consensus trained model was 0.33, while the consensus model was 0.57.}
    \label{fig:concordance_vs_discordance}
\end{figure}

Diagnosing melanocytic cases is challenging. Although some specimens (such as ones diagnosed as Compound Nevi) clearly exhibit very low risk and others (such as invasive melanoma) exhibit very high risk of progressing into life threatening conditions, reproducible stratification in the middle of the morphological spectrum has historically proved difficult\cite{corona1996interobserver, elmore2017pathologists}. The results were derived with our PDLS trained and evaluated on \textit{consensus} data: data for which the ground truth melanocytic specimen diagnostic categories were agreed upon by multiple experts. To understand the effect of consensus on training deep learning models, we performed an ablation study by training two hierarchical models. Both models used \emph{all} non-melanocytic specimens available in the training set. The first model was trained only including melanocytic specimens for which consensus was obtained under the diagnostic categories of MPATH I/II, MPATH III, or MPATH IV/V. The other model was trained by also including \textit{non-consensus} data: melanocytic specimens whose diagnostic category was not agreed upon by the experts. To facilitate a fair comparison, we reserved separate validation sets for both model versions and a common \textit{consensus} test set derived from the Reference Lab. The sensitivities of our PDLS to different classes on both \textit{consensus} and \textit{non-consensus} data are shown in Figure \ref{fig:concordance_vs_discordance}, where a a clear improvement is shown in the sensitivity to the Melanocytic class of over $40\%$ for melanocytic specimens that are annotated with consensus labels over ones that are not; this primarily manifested from a reduction in false positive Melanocytic Suspect classifications.

\begin{figure}
     \includegraphics[scale=0.35]{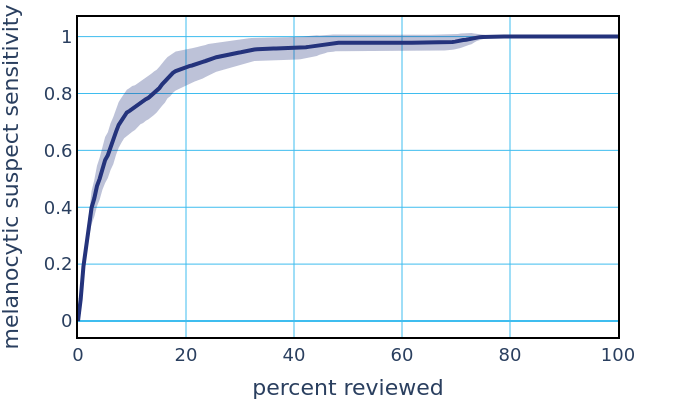}
     \centering
    \caption{Mean and standard deviation sensitivity to melanoma vs \% reviewed for 1,000 simulated sequentially-accessioned datasets, drawn from Reference lab confidence scores. In the clinic, 95\% of melanoma suspect cases are detected within the first 30\% of cases, when ordered by melanoma suspect model confidence.}
    \label{fig:recall_vs_confidence}
\end{figure}


\section{Discussion}
\label{sec:discussion}


In this work, we demonstrated the performance of a PDLS capable of automatically sorting and triaging skin specimens with high sensitvity to Melanocytic Suspect cases prior to review by a pathologist.  Existing methods (such as \cite{mypath}) provide diagnostically-relevant information on a potential melanoma specimen only \textit{after} a pathologist has reviewed the specimen and classified it as a Melanocytic Suspect lesion. The PDLS's ability to classify suspected melanoma prior to pathologist review could substantially reduce diagnostic turnaround time for melanoma by not only allowing timely review and expediting the ordering of additional tests or stains,  but also ensuring that suspected melanoma cases are routed directly to subspecialists.  The potential clinical impact of a PDLS with these capabilities is underscored by the fact that early melanoma detection is correlated with improved patient outcomes \cite{matthewsmel}.

As the PDLS was optimized to maximize melanoma sensitivty, we investigated the performance of our PDLS as a simple Melanocytic Suspect binary classifier. We envisioned a scenario in which a pathologist's work list of specimens is sorted by the system's confidence (in descending order) in the upstream classifier's suspect melanocytic classification. Figure \ref{fig:recall_vs_confidence} demonstrates the resulting sensitivity to the Melanocytic Suspect class against the percentage of total specimens that a pathologist would have to review in this scheme in order to achieve that sensitivity.  In this dataset, we show that a pathologist would only need between 30\% and 60\% of the caseload to address \emph{all} melanoma specimens.

Diagnostic classification of melanocytic lesions remains challenging. There is known lack of consensus among pathologists, and a disturbing lack of intra-pathologist concordance over time \cite{onega2018accuracy} was recently reported. Training with consensus data resulted in improved performance seen in classifications \emph{excluding} Melanocytic Suspect, which has the highest pathologist discordance rates. Since pathologists tend to cautiously diagnose a benign lesion as malignant, our PDLS learned the same bias in absence of consensus. By training on consensus of multiple dermatopathologists, our PDLS may have the unique ability to learn a more consistent feature representation of melanoma and aid in flagging misdiagnosis. While the system is highly sensitive to melanoma ($~84\%$ correctly detected as Intermediate or High Risk in the Reference Lab Test set) there are a large number of false positives (~$2.7\%$ of sequentially-accessioned specimens in the reference lab were predicted to be suspect) classified as suspect. It may therefore be possible to flag initial diagnoses discordant with the PDLS classification of highly confident predictions for review in order to lower the false positive rate. 

The PDLS system also enables other automated pathology workflows in addition to triage and prioritization of suspected melanoma cases. Sorting and triaging specimens into other classifications such as Basaloid could allow the majority of less complicated cases (such as basal cell carcinoma) to be directly assigned to general pathologists, or to dermatologists who routinely sign out such cases.

Relevant to any PDLS designed for clinical use is how well its performance generalizes to sites on which the system was not trained. Performance on the Validation Labs after calibration (as shown in Figure \ref{fig:results}) was in many cases close to that of the Reference Lab. Further improvement is needed, however, in order to avoid the need for fine-tuning and improve model generalizability of previously trained models in new laboratory settings.

{\small
\bibliographystyle{unsrt}
\bibliographystyle{ieee_fullname}
\bibliography{bibliography.bib}
}

\end{document}